\documentclass[%
 prl,
 twocolumn,
 longbibliography,
 amsmath,amssymb,
 aps,
 showkeys,
]{revtex4-1}

\usepackage{graphicx}
\usepackage{dcolumn}
\usepackage{bm}
\usepackage{hyperref}

\begin{document}

\preprint{APS/123-QED}

\title{How does a hyperuniform fluid freeze?}

\author{Yusheng Lei}
\author{Ran Ni}
\email{r.ni@ntu.edu.sg}
\affiliation{School of Chemistry, Chemical Engineering and Biotechnology, Nanyang Technological University, 62 Nanyang Drive, 637459, Singapore}%

\date{\today}

\begin{abstract}
All phase transitions can be categorised into two different types: continuous and discontinuous phase transitions. Discontinuous phase transitions are normally accompanied with significant structural changes, and nearly all of them have the kinetic pathway of nucleation and growth, if the system does not suffer from glassy dynamics. Here, in a system of barrier-controlled reactive particles, we find that the discontinuous freezing transition of a non-equilibrium hyperuniform fluid into an absorbing state does not have the kinetic pathway of nucleation and growth, and the transition is triggered by long wavelength fluctuations. The transition rate decreases with increasing the system size, which suggests that the metastable hyperuniform fluid is kinetically stable in an infinitely large system. This challenges the common understanding of metastability in discontinuous phase transitions. Moreover, we find that the ``metastable yet kinetically stable'' hyperuniform fluid features a new scaling in the structure factor $S(k \rightarrow 0) \sim k^{1.2}$ in 2D, which is the third dynamic hyperuniform state in addition to the critical hyperuniform state with $S(k \rightarrow 0) \sim k^{0.45}$ and the non-equilibrium hyperuniform fluid with $S(k \rightarrow 0) \sim k^{2}$.
\end{abstract}

\keywords{non-equilibrium hyperuniform fluid, discontinuous phase transition, long wavelength fluctuation, metastable yet kinetically stable}
\maketitle

\twocolumngrid
The concept of hyperuniformity was introduced by Torquato and Stillinger in 2003~\cite{torquato2003local}, in which a hyperuniform structure is defined if the structure factor $S(|\mathbf{k}| \rightarrow 0 ) = 0$ with $\mathbf{k}$ the wavevector. Besides the ``ordered'' hyperuniform structures, like crystals and quasicrystals~\cite{ouguz2017hyperuniformity}, over the past decades, a number of disordered hyperuniform structures have been found in various systems, including perturbed lattices~\cite{kim2018effect}, perfect glasses~\cite{zhang2016perfect}, jammed structures \cite{donev2005unexpected, ricouvier2017optimizing}, avian photoreceptor patterns~\cite{jiao2014avian}, biological tissues~\cite{zheng2020hyperuniformity}, early universe fluctuations~\cite{gabrielli2002glass}, etc. These disordered hyperuniform structures have shown even better properties than the ordered hyperuniform structures like isotropic photonic bandgaps opened at low dielectric contrast~\cite{florescu2009designer,man2013isotropic,manoe2013} and abnormal transparency~\cite{transp2008,leseur2016high}, which suggests a new direction in design and fabrication of disordered hyperuniform functional materials.
Recently, emergent dynamic hyperuniform states were also found in non-equilibrium systems theoretically~\cite{hexner2017noise,lei2019hydrodynamics},  numerically~\cite{Mitra_2021, hexner2015hyperuniformity,hexner2017noise,tjhung2015hyperuniform,lei2019nonequilibrium,lei2019hydrodynamics, wang2018hyperuniformity,  oppenheimer2022hyperuniformity} and experimentally~\cite{zhang2022hyperuniform,huang2021circular,chaikinprl2020,weijsprl2015}, and most of them can be categorised into the critical hyperuniform state and the non-equilibrium hyperuniform fluid. It is found that the non-equilibrium hyperuniform fluid originates from the interplay between the reciprocal active excitation between particles and the frictional dissipation of the background solvent. With decreasing the active excitation or increasing the friction of the solvent, the system undergoes an absorbing transition into an immobile state~\cite{lei2019hydrodynamics}. While most absorbing transitions are continuous phase transitions, it was recently found that in systems of barrier-controlled reactive particles, by increasing the reaction barrier, the absorbing transition can be discontinuous~\cite{lei2021barrier}. Here, using mean field theory and computer simulation, we investigate the kinetic pathway for the non-equilibrium hyperuniform fluid of barrier-controlled reactive particles undergoing the discontinuous phase transition into an  absorbing state. Intriguingly, we find that the discontinuous absorbing transition of non-equilibrium hyperuniform fluid does not have the kinetic pathway of nucleation and growth, which suggests that the metastable hyperuniform fluid is kinetically stable in the thermodynamic limit with a new hyperuniform scaling in the structure factor $S( k \rightarrow 0 ) \sim k^{1.2}$.

\begin{figure*}[htb]
    \centering
    \includegraphics[width=0.95\textwidth]{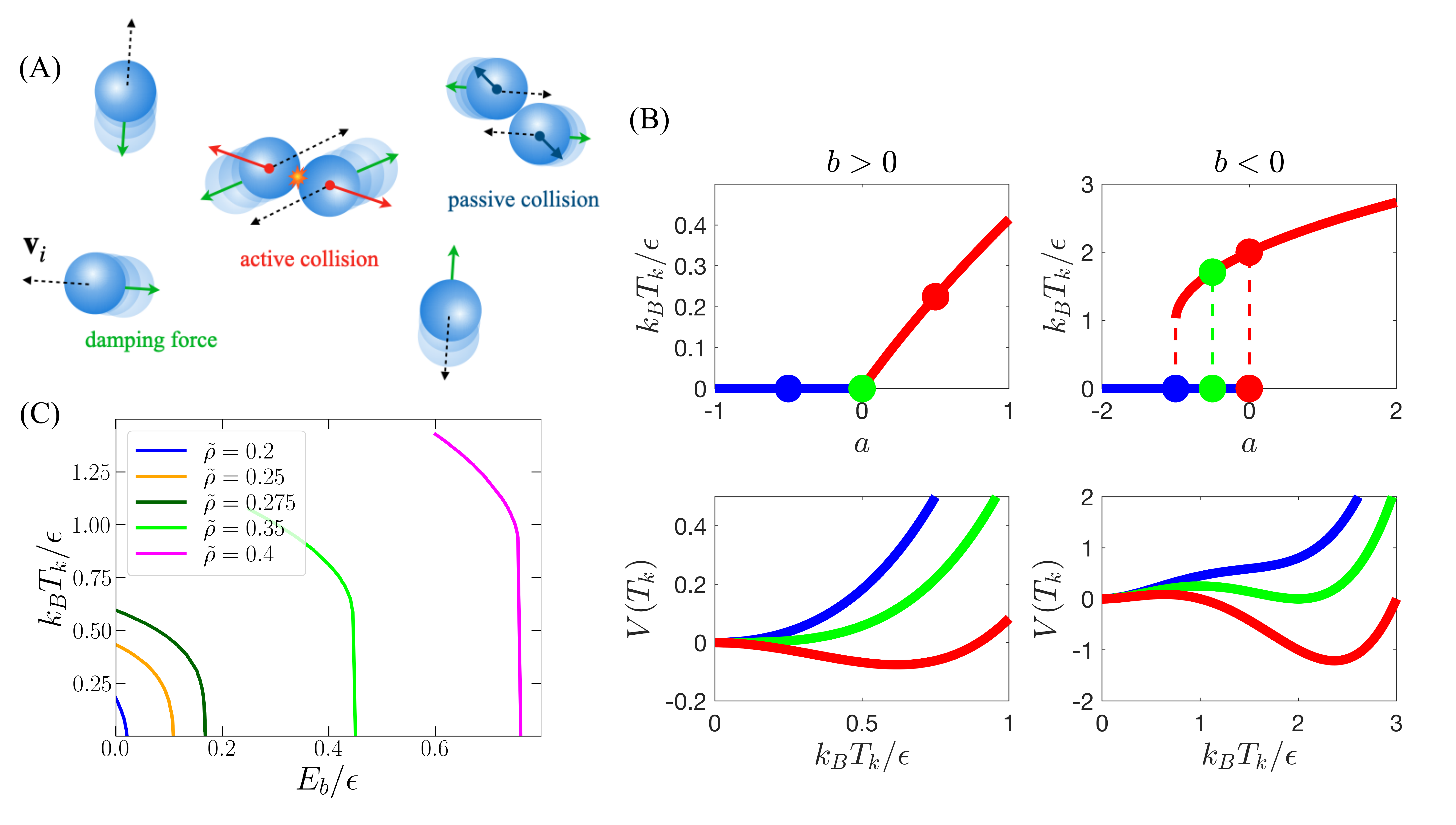}
    \caption{{\bf Barrier-controlled reactive particles.} (A): Schematic of the barrier-controlled reactive hard spheres. (B): Schematic of the mean field prediction on the phase behavior of  barrier-controlled reactive particles, in which depending on the parameter $b$, $T_k$ changes with parameter $a$ either continuously or discontinuously to zero. Here $V(T_k)$ is the effective potential constructed in the mean field theory. (C): Steady-state kinetic temperature $T_k$ as a function of reaction barrier $E_b$ for systems of $N=10000$ particles at various density $\tilde{\rho}$ with $l_d=\sqrt{5}\sigma$.}
    \label{Fig.1}
\end{figure*}

\section*{Results}
\subsection*{Model}
We consider a generalized reactive particle system in $d$-dimension consisting of $N$ hard spheres with mass $m$, diameter $\sigma$ and random initial velocities $\mathbf{v}$~\cite{lei2021barrier} as shown in Fig.~\ref{Fig.1}A, and we focus on $d=2$. Particles can undergo active or passive collisions depending on their colliding velocities. For two particles $(\mathbf{r}_i,\mathbf{v}_i)$ and $(\mathbf{r}_j,\mathbf{v}_j)$ colliding at time $t$, if the relative kinetic energy between them is larger than the reaction barrier $E_b$, i.e., $\frac{1}{2}m(\Delta v_{i,j}^\perp)^2>E_b$, they undergo an active collision, and an extra energy $\epsilon$ is injected reciprocally to the kinetic energy of the two particles in the colliding direction, otherwise the two particles undergo an elastic passive collision. Here $\Delta v_{i,j}^\perp$ is the relative velocity along the center-to-center direction, and $\Delta v_{i,j}^\perp=\Delta \mathbf{v}_{i,j}\cdot \frac{\Delta \mathbf{r}_{i,j}}{\sigma}$ with $\Delta \mathbf{r}_{i,j}=\mathbf{r}_j-\mathbf{r}_i$.
The equation of motion for particle $i$ between two consecutive collisions is described by the underdamped Langevin equation
\begin{equation}
    m \frac{d \mathbf{v}_i(t)}{d t}=-\gamma \mathbf{v}_i(t)
\end{equation}
with the friction coefficient $\gamma$. 
The system is simulated using an event-driven algorithm~\cite{lei2019hydrodynamics,lei2021barrier}.
The dimensionless particle density of the system is $\tilde{\rho}\equiv N \sigma^d/V$, where $V=L^d$ is the volume of the system, and $L$  is the box length with periodic boundary conditions in all directions. The dissipation length $l_d\equiv\sqrt{m\epsilon}/\gamma$, and $l_d$ is set as $\sqrt{5}\sigma$ in this work. Here the time unit is $\tau_0\equiv\sigma/v_0$ with the typical excitation speed defined as $v_0\equiv\sqrt{\epsilon/m}$.

\begin{figure*}[htb]
    \centering
    \includegraphics[width=0.95\textwidth]{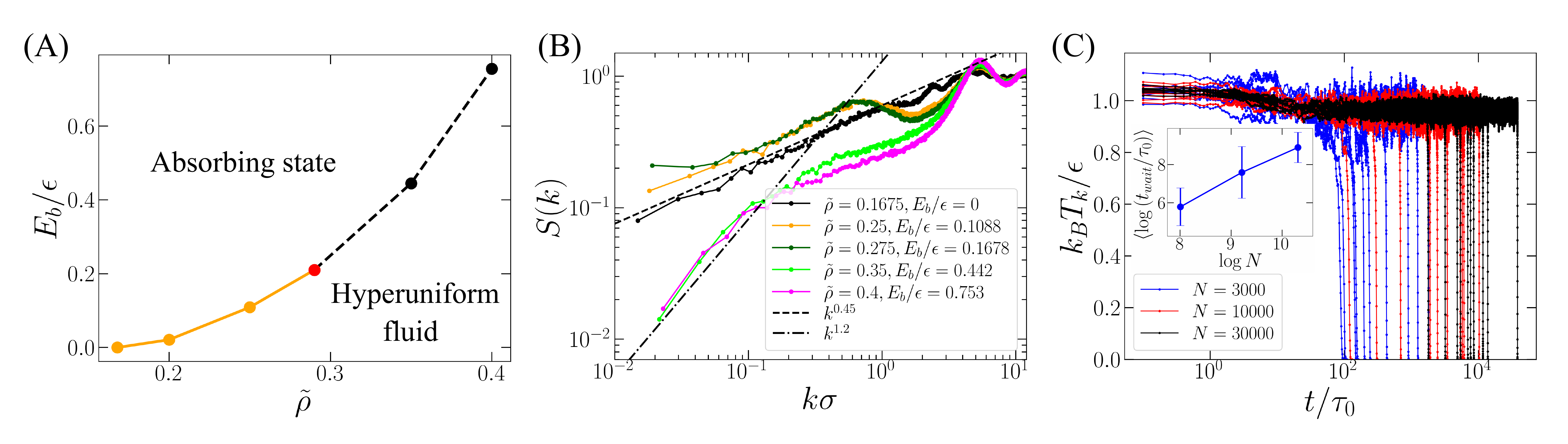}
    \caption{{\bf Phase behaviour of barrier-controlled reactive hard spheres.} (A): Phase diagram of barrier-controlled reactive hard spheres, where the orange curve indicates the continuous absorbing transition, the red dot indicates the tricritical point, and the black dashed curve indicates the stability limit of hyperuniform fluid in the discontinuous absorbing transition. (B): The structure factor near the critical point of continuous phase transition ($\tilde{\rho}=0.1675,\,0.25$), the tricritical point $\tilde{\rho}_{\rm tri}=0.275,\,E_{b, \rm tri}/\epsilon=0.1678$, and the stability limit of the hyperuniform fluid in the discontinuous phase transition ($\tilde{\rho}=0.35,\,0.4$). Here the dashed and dotted dashed lines indicate two hyperuniform scaling $S(k \rightarrow 0 ) \sim k^{0.45}$ and $\sim k^{1.2}$, respectively. (C): The time evolution of the kinetic temperature $T_k$ in the metastable hyperuniform fluid of different sizes at $\tilde{\rho}=0.4,\,E_b/\epsilon=0.753$, in which for each system size, we perform 10 independent simulations. Inset: $\langle \log (t_{\rm wait}/\tau_0) \rangle$ as a function of the system size $N$, where errorbars are standard deviations.}
    \label{Fig.2}
\end{figure*}

\subsection*{Mean field theory}
We first formulate a qualitative mean field theory to describe the phase behaviour of the system, which is essentially a driven-dissipative system~\cite{lei2021barrier}. At the mean field level, the power of energy per particle for the driven-flow by active collisions and the dissipative-flow by the solvent can be written as $W_{\text {driv}}=f_a \epsilon$ and $W_{\mathrm{disp}}= \gamma\overline{v^2}$, respectively. We define the kinetic temperature of the system $T_k\equiv m \langle \overline{v^2} \rangle /dk_B$, with which we can rewrite $W_{\mathrm{disp}}= \gamma d k_B T_k /m$.
$f_a$ is the average active collision frequency per particle, which can be approximated as $f_a \simeq x_a \bar{v}_a/(2l_r)$, where $x_a$, $\bar{v}_a$, and $l_r$ are the fraction, average speed, and mean free path of active particles, respectively. At low density, we assume $l_r \simeq \sigma/\tilde{\rho}_r$ with $\tilde{\rho}_r$ the density of effective reactants, i.e., the average density of particles that can be activated at $T_k$. Therefore, the mean field dynamic equation for $T_k$ can be written as
\begin{equation}\label{eq:mf}
    \frac{\partial (dk_B T_k)}{\partial t} = W_{\text {driv}}-W_{\mathrm{disp}} =\frac{x_a \tilde{\rho}_r \bar{v}_a \epsilon}{2 \sigma}- \frac{\gamma dk_B T_k}{m} .
\end{equation}
Increasing $x_a$ encourages more collisions between active particles to increase $\bar{v}_a$, and as a first-order approximation, we assume $\bar{v}_a \simeq (1+Ax_a)v_0$, with $A$ a positive constant. Then we have $d k_B T_k \simeq x_a m \bar{v}_a^2 \simeq (x_a + 2A x_a^2) \epsilon$. Accordingly, as a first-order approximation, we have $x_a \simeq dk_BT_k/\epsilon-2 Ad^2k_B^2 T_k^2/\epsilon^2$. Moreover, increasing the reaction barrier $E_b$ decreases the effective reactant density $\tilde{\rho}_r$, and as a first-order approximation, we assume $\tilde{\rho}_r/\tilde{\rho} \simeq 1 - B(1-x_a)E_b/\epsilon$ with $B$ a positive constant. Thus, by keeping only the first three leading terms, Eq.~\eqref{eq:mf} can be written as
\begin{equation}\label{eq:cubicmf}
\frac{\partial  T_k}{\partial t}=a T_k-b T_k^2-c T_k^3,
\end{equation}
where $a=\frac{\tilde{\rho}}{2\tau_0}[(1-B E_b/\epsilon)-(\tau_0/\tau_d)]$, $b=-\frac{dk_B\tilde{\rho}}{2 \tau_0 \epsilon^2 }[(1+A) B E_b - A\epsilon]$, $c=\frac{d^2k_B^2}{2\tau_0 \epsilon^3}\tilde{\rho}[4 \epsilon A^2+(3 A-4 A^2) B E_b]$, and $\tau_d=m/\gamma$.
Eq.~\ref{eq:cubicmf} with $b<0$ is the simplest equation that was used to investigate catastrophic shifts at a deterministic level~\cite{villa2015eluding}.
The right hand side of Eq.~\ref{eq:cubicmf} can be seen as the gradient of an effective potential $V(T_k)$, and $\partial T_k / \partial t = - \partial V / \partial T_k$, in which the sign of the parameter $b$ controls the nature of the transition as shown in Fig.~\ref{Fig.1}B.
For $b > 0$, one can see that $V(T_k)$ only has one minimum suggesting that the transition from active state ($T_k > 0$) to the absorbing state ($T_k = 0$) is a continuous transition. When $b < 0$, $V(T_k)$ features two local minima, which implies that with decreasing $a$, the system undergoes a discontinuous transition from the active state to the active-absorbing bistable region and then to the absorbing state. This can be also understood as that the existance of reaction barrier delays the dynamic phase transition but also increases the cooperativity during the transition, which sharpens the transition. With a high enough reaction barrier, the absorbing transition becomes discontinuous~\cite{lei2021barrier}.
$b = 0$ is the tricritical point that separates the continuous and discontinuous phase transitions~\cite{lei2021barrier}.

To validate the theoretical prediction on the phase behaviour, we perform event-driven simulations for a 2D system of $N=10000$ reactive hard spheres at various densities $\tilde{\rho}$ and reaction barriers $E_b$. All simulations start from random configurations and velocities of sufficiently large initial kinetic temperature, and we use the kinetic temperature $T_k$ as the order parameter of the system. As shown in Fig.~\ref{Fig.1}C, for low density systems, $T_k$ decreases smoothly to zero with increasing $E_b$, and with increasing the density of the system, the absorbing transition occurs at higher $E_b$ with the transition becoming sharper. With the finite size scaling of the dynamic evolution of the system, we confirm that at $\tilde{\rho}_{\rm tri} = 0.275$ and $E_{b,\rm tri}/\epsilon = 0.1678$, the kinetic temperature decays as $T_k \sim t^{-0.37}$ before reaching the saturated value, which signatures a tricritical point (see SI). For systems at lower densities, at the transition point, $T_k \sim t^{-0.54}$, which belongs to the conserved directed percolation universality class like the conventional absorbing transition (see SI). For systems at higher densities, we find that the absorbing transition is discontinuous, and these qualitatively confirm the prediction from the mean field theory.

\begin{figure*}[htb]
    \centering
    \includegraphics[width=0.95\textwidth]{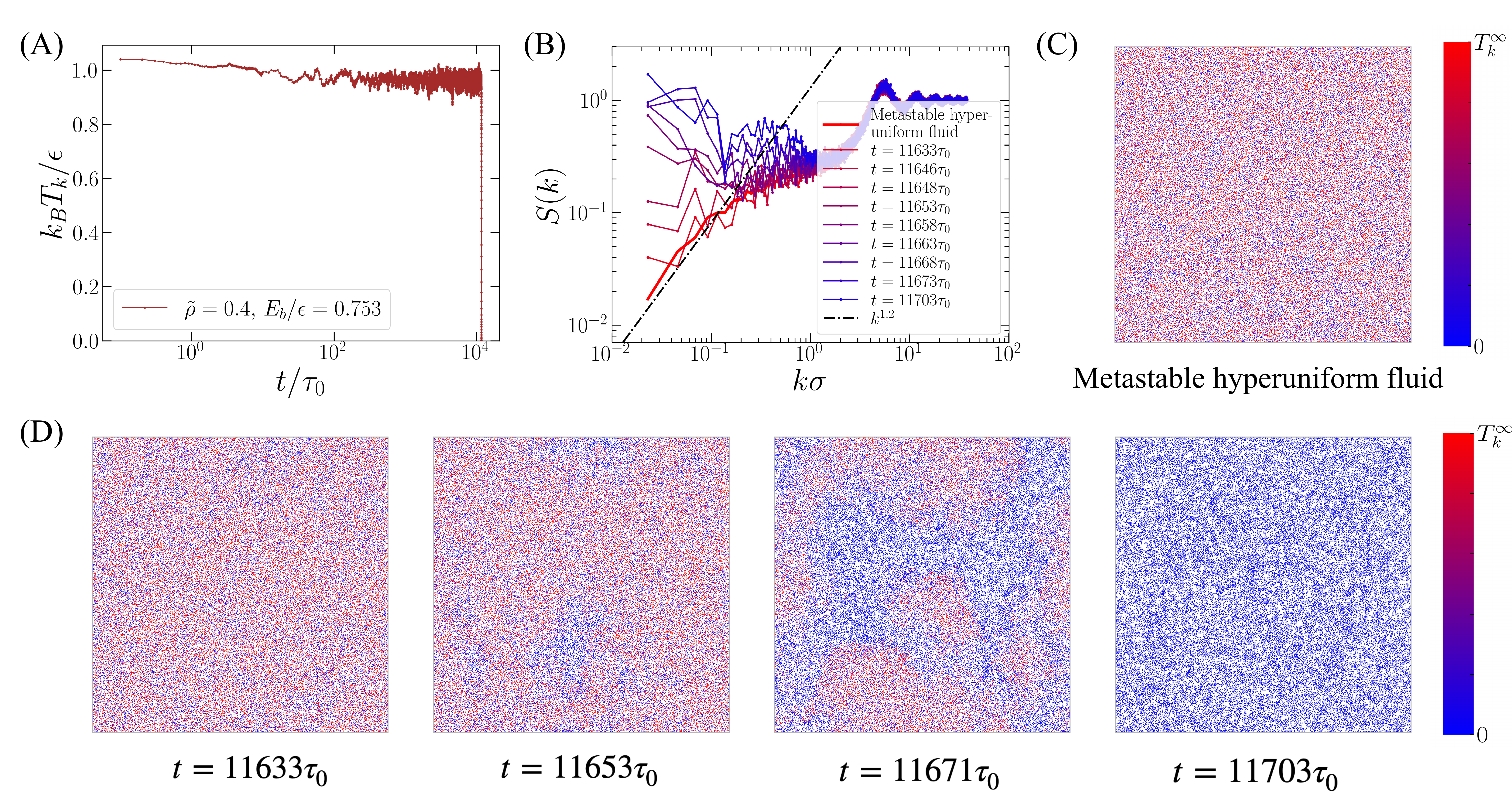}
    \caption{{\bf Kinetic pathway of the discontinuous absorbing transition.} (A): The time evolution of the kinetic temperature of the system of $N=30000$ reactive particles at $\tilde{\rho} = 0.4$ and $E_b/\epsilon = 0.753$. (B): The structure factor for system at various times during the discontinuous phase transition from a metastable hyperuniform fluid to an absorbing state, where the dotted dashed lines indicates the hyperuniform scaling of $S(k \rightarrow 0 ) \sim k^{1.2}$. (C): A typical snapshot of the metastable hyperuniform fluid. (D): The snapshots of the system along a typical dynamic trajectory of the discontinuous absorbing transition. In (C) and (D), the particles are color coded based on their kinetic temperature, and $T_{k}^{\infty}$ is the kinetic temperature of the metastable hyperuniform fluid.}
    \label{Fig.3}
\end{figure*}

\subsection*{Phase diagram}
The phase diagram of barrier-controlled reactive particles is summarized in Fig.~\ref{Fig.2}A, which features both a continuous and a discontinuous absorbing transition separated by a tricritical point located at around $\tilde{\rho}_{\rm tri} = 0.275$ and $E_{b,\rm tri}/\epsilon = 0.1678$. We further check the structure of the system at the transition points. As shown in Fig.~\ref{Fig.2}B, at the continuous absorbing transition, $S(k \rightarrow 0 ) \sim k^{0.45}$. At the tricritical point, the structure factor follows the same scaling as the conventional continuous absorbing transition, although the dynamic relaxation is different. For systems at higher densities that undergo discontinuous absorbing transition, we find that the structure factor of the metastable hyperuniform fluid $S(k \rightarrow 0 ) \sim k^{1.2}$, which is a new hyperuniform scaling in non-equilibrium dynamic systems. Based on the obtained phase diagram, we investigate the kinetic pathway of the discontinuous absorbing transition. We simulate the metastable hyperuniform fluids of different system size starting from random configurations, and for each system, we perform 10 independent simulations, in which the time evolution of the kinetic temperature $T_k$ is shown in Fig.~\ref{Fig.2}C.
One can see that $T_k$ reaches a plateau around $\epsilon/k_B$ quickly, and remains constant for a period of time $t_{\rm wait}$ before suddenly drops to zero, which is a signature of discontinuous phase transition~\cite{ludopre2016,ludopre2017}. It is known that the discontinuous phase transition is a barrier-crossing process, of which the barrier height is related to $\log (t_{\rm wait})$. In the inset of Fig.~\ref{Fig.2}C, we plot $\langle \log (t_{\rm wait}/\tau_0) \rangle$ as a function of system size $N$.
Intriguingly, we find that $\langle \log (t_{\rm wait}/\tau_0) \rangle$ increases monotonically with $N$, which suggests that the kinetic pathway is not nucleation and growth. Because if the kinetic pathway is nucleation and growth, one shall have a well defined nucleation rate, which is the number of critical nuclei found in a system of unit volume within a unit time period. With increasing the system size  $N$, the probability of finding a critical nucleus in the system within a unit time increases, which should make the waiting time $t_{\rm wait}$ in the metastable state shorter, while in Fig.~\ref{Fig.2}C, $t_{\rm wait}$ counterintuitively increases with $N$.

\subsection*{Kinetic pathway of phase transition}
To understand the physics of the discontinuous absorbing transition, we investigate a typical dynamic trajectory of phase transforming hyperuniform fluid of $\tilde{\rho} = 0.4$ at $E_b/\epsilon = 0.753$, of which the time evolution of $T_k$ is shown in Fig.~\ref{Fig.3}A. One can see that the kinetic temperature of the system reaches the metastable steady state with $k_B T_k/\epsilon \approx 1$ within about $10^2 \tau_0$ and drops abruptly to zero at around $t_{\rm wait} \approx 10^4 \tau_0$. A movie of the phase transition can be found in SI.
In Fig.~\ref{Fig.3}B, we plot the structure factor of the system along the phase transition in comparison with the starting metastable hyperuniform fluid. One can see that at the beginning of the phase transition, e.g., $t = 11633 \tau_0$, the structure factor starts to deviate from the metastable hyperuniform fluid in the smallest $k$ region corresponding to the wavelength of system size, while the local structure, i.e., $1 < k \sigma  < 10$,  remains intact. The corresponding snapshots of the systems along the transition are shown in Fig.~\ref{Fig.3}D, and one can see that compared with the typical snapshot of the metastable hyperuniform fluid in Fig.~\ref{Fig.3}C, there is hardly any visible difference at the beginning of the transition. This is due to the fact that the structural difference appears first in the small $k$ region corresponding to the long range correlation, and the structural difference only becomes visible when the deviation propagates to the intermediate lengthscale, e.g., $t = 11653 \tau_0$ in Fig.~\ref{Fig.3}D, where multiple clusters of immobile clusters appear simultaneously like spinodal decomposition~\cite{niprl2010}. Afterwards, the clusters of immobile particles percolate, and the system transforms into an absorbing state. This implies that the phase transition is triggered by long wavelength fluctuations, and explains why $t_{\rm wait}$ increases with $N$. Because the probability of having a fluctuation of the wavelength of the system size decreases with increasing the system size. This suggests that in the thermodynamic limit, the rate of discontinuous absorbing transition is zero.
 
\section*{Conclusion and Discussions}
In conclusion, using mean field theory and computer simulation, we have investigated a system of barrier-controlled reactive particles forming a non-equilibrium hyperuniform fluid, which can undergo either continuous or discontinuous phase transitions to an absorbing state depending on the reaction barrier. Intriguingly, we find that the discontinuous freezing of the metastable hyperuniform fluid into the absorbing state does not have the kinetic pathway of nucleation and growth, and the transition rate decreases with increasing the system size. By checking the structural change along the phase transformation, we find that the discontinuous absorbing transition is triggered by long wavelength fluctuations, of which the probability decreases with increasing the system size. The absence of the kinetic pathway of nucleation and growth in the metastable hyperuniform fluid can be understood as follows. To have a kinetic pathway of nucleation and growth for any discontinuous phase transition, there must be a coexisting region of the two phases, in which local structural difference is necessary for creating an interface. Here, in the discontinuous absorbing transition, the structural difference between the metastable hyperuniform fluid and the absorbing state only exists at small $k$ corresponding to long range correlations, and there is no local structural difference between them near $k \approx 2\pi/\sigma$ (Fig.~\ref{Fig.3}B). Therefore, one cannot have the coexistence of the hyperuniform fluid and the absorbing state in one system, which suggests that there is no kinetic pathway of nucleation and growth in the discontinuous absorbing transition from the metastable hyperuniform fluid.
This suggests that the metastable hyperuniform fluid is kinetically stable in an infinitely large system, which challenges the common understanding of metastability in discontinuous phase transitions. Lastly, the ``metastable yet kinetically stable'' hyperuniform fluid features a new hyperuniform scaling $S(k\rightarrow 0) \sim k^{1.2}$, which is different from the other dynamic hyperuniform states, i.e., the critical hyperuniform state and the non-equlibrium hyperuniform fluid.

\begin{acknowledgments}
\textbf{Acknowledgments:}
This work has been supported by the Singapore Ministry of Education through the Academic Research Fund
MOE2019-T2-2-010. \textbf{Author contributions:} R.N. conceived and directed the research; Y.L. performed the research; Y.L. and R.N. analysed the data and wrote the manuscript. \textbf{Competing interests:} The authors declare that they have no competing interests. \textbf{Data and materials availability:} All data needed to evaluate the conclusions in the paper are presented in the paper and/or the Supplementary Materials. Additional data related to this paper may be requested from the authors.
\end{acknowledgments}

\bibliography{paper}

\begin{thebibliography}{31}%
\makeatletter
\providecommand \@ifxundefined [1]{%
 \@ifx{#1\undefined}
}%
\providecommand \@ifnum [1]{%
 \ifnum #1\expandafter \@firstoftwo
 \else \expandafter \@secondoftwo
 \fi
}%
\providecommand \@ifx [1]{%
 \ifx #1\expandafter \@firstoftwo
 \else \expandafter \@secondoftwo
 \fi
}%
\providecommand \natexlab [1]{#1}%
\providecommand \enquote  [1]{``#1''}%
\providecommand \bibnamefont  [1]{#1}%
\providecommand \bibfnamefont [1]{#1}%
\providecommand \citenamefont [1]{#1}%
\providecommand \href@noop [0]{\@secondoftwo}%
\providecommand \href [0]{\begingroup \@sanitize@url \@href}%
\providecommand \@href[1]{\@@startlink{#1}\@@href}%
\providecommand \@@href[1]{\endgroup#1\@@endlink}%
\providecommand \@sanitize@url [0]{\catcode `\\12\catcode `\$12\catcode
  `\&12\catcode `\#12\catcode `\^12\catcode `\_12\catcode `\%12\relax}%
\providecommand \@@startlink[1]{}%
\providecommand \@@endlink[0]{}%
\providecommand \url  [0]{\begingroup\@sanitize@url \@url }%
\providecommand \@url [1]{\endgroup\@href {#1}{\urlprefix }}%
\providecommand \urlprefix  [0]{URL }%
\providecommand \Eprint [0]{\href }%
\providecommand \doibase [0]{http://dx.doi.org/}%
\providecommand \selectlanguage [0]{\@gobble}%
\providecommand \bibinfo  [0]{\@secondoftwo}%
\providecommand \bibfield  [0]{\@secondoftwo}%
\providecommand \translation [1]{[#1]}%
\providecommand \BibitemOpen [0]{}%
\providecommand \bibitemStop [0]{}%
\providecommand \bibitemNoStop [0]{.\EOS\space}%
\providecommand \EOS [0]{\spacefactor3000\relax}%
\providecommand \BibitemShut  [1]{\csname bibitem#1\endcsname}%
\let\auto@bib@innerbib\@empty
\bibitem [{\citenamefont {Torquato}\ and\ \citenamefont
  {Stillinger}(2003)}]{torquato2003local}%
  \BibitemOpen
  \bibfield  {author} {\bibinfo {author} {\bibfnamefont {Salvatore}\
  \bibnamefont {Torquato}}\ and\ \bibinfo {author} {\bibfnamefont {Frank~H}\
  \bibnamefont {Stillinger}},\ }\bibfield  {title} {\enquote {\bibinfo {title}
  {Local density fluctuations, hyperuniformity, and order metrics},}\
  }\href@noop {} {\bibfield  {journal} {\bibinfo  {journal} {Phys. Rev. E}\
  }\textbf {\bibinfo {volume} {68}},\ \bibinfo {pages} {041113} (\bibinfo
  {year} {2003})}\BibitemShut {NoStop}%
\bibitem [{\citenamefont {O{\u{g}}uz}\ \emph {et~al.}(2017)\citenamefont
  {O{\u{g}}uz}, \citenamefont {Socolar}, \citenamefont {Steinhardt},\ and\
  \citenamefont {Torquato}}]{ouguz2017hyperuniformity}%
  \BibitemOpen
  \bibfield  {author} {\bibinfo {author} {\bibfnamefont {Erdal~C}\ \bibnamefont
  {O{\u{g}}uz}}, \bibinfo {author} {\bibfnamefont {Joshua~ES}\ \bibnamefont
  {Socolar}}, \bibinfo {author} {\bibfnamefont {Paul~J}\ \bibnamefont
  {Steinhardt}}, \ and\ \bibinfo {author} {\bibfnamefont {Salvatore}\
  \bibnamefont {Torquato}},\ }\bibfield  {title} {\enquote {\bibinfo {title}
  {Hyperuniformity of quasicrystals},}\ }\href@noop {} {\bibfield  {journal}
  {\bibinfo  {journal} {Phys. Rev. B}\ }\textbf {\bibinfo {volume} {95}},\
  \bibinfo {pages} {054119} (\bibinfo {year} {2017})}\BibitemShut {NoStop}%
\bibitem [{\citenamefont {Kim}\ and\ \citenamefont
  {Torquato}(2018)}]{kim2018effect}%
  \BibitemOpen
  \bibfield  {author} {\bibinfo {author} {\bibfnamefont {Jaeuk}\ \bibnamefont
  {Kim}}\ and\ \bibinfo {author} {\bibfnamefont {Salvatore}\ \bibnamefont
  {Torquato}},\ }\bibfield  {title} {\enquote {\bibinfo {title} {Effect of
  imperfections on the hyperuniformity of many-body systems},}\ }\href@noop {}
  {\bibfield  {journal} {\bibinfo  {journal} {Phys. Rev. B}\ }\textbf {\bibinfo
  {volume} {97}},\ \bibinfo {pages} {054105} (\bibinfo {year}
  {2018})}\BibitemShut {NoStop}%
\bibitem [{\citenamefont {Zhang}\ \emph {et~al.}(2016)\citenamefont {Zhang},
  \citenamefont {Stillinger},\ and\ \citenamefont
  {Torquato}}]{zhang2016perfect}%
  \BibitemOpen
  \bibfield  {author} {\bibinfo {author} {\bibfnamefont {Ge}~\bibnamefont
  {Zhang}}, \bibinfo {author} {\bibfnamefont {Frank~H}\ \bibnamefont
  {Stillinger}}, \ and\ \bibinfo {author} {\bibfnamefont {Salvatore}\
  \bibnamefont {Torquato}},\ }\bibfield  {title} {\enquote {\bibinfo {title}
  {The perfect glass paradigm: Disordered hyperuniform glasses down to absolute
  zero},}\ }\href@noop {} {\bibfield  {journal} {\bibinfo  {journal} {Sci.
  Rep.}\ }\textbf {\bibinfo {volume} {6}},\ \bibinfo {pages} {1--12} (\bibinfo
  {year} {2016})}\BibitemShut {NoStop}%
\bibitem [{\citenamefont {Donev}\ \emph {et~al.}(2005)\citenamefont {Donev},
  \citenamefont {Stillinger},\ and\ \citenamefont
  {Torquato}}]{donev2005unexpected}%
  \BibitemOpen
  \bibfield  {author} {\bibinfo {author} {\bibfnamefont {Aleksandar}\
  \bibnamefont {Donev}}, \bibinfo {author} {\bibfnamefont {Frank~H}\
  \bibnamefont {Stillinger}}, \ and\ \bibinfo {author} {\bibfnamefont
  {Salvatore}\ \bibnamefont {Torquato}},\ }\bibfield  {title} {\enquote
  {\bibinfo {title} {Unexpected density fluctuations in jammed disordered
  sphere packings},}\ }\href@noop {} {\bibfield  {journal} {\bibinfo  {journal}
  {Phys. Rev. Lett.}\ }\textbf {\bibinfo {volume} {95}},\ \bibinfo {pages}
  {090604} (\bibinfo {year} {2005})}\BibitemShut {NoStop}%
\bibitem [{\citenamefont {Ricouvier}\ \emph {et~al.}(2017)\citenamefont
  {Ricouvier}, \citenamefont {Pierrat}, \citenamefont {Carminati},
  \citenamefont {Tabeling},\ and\ \citenamefont
  {Yazhgur}}]{ricouvier2017optimizing}%
  \BibitemOpen
  \bibfield  {author} {\bibinfo {author} {\bibfnamefont {Joshua}\ \bibnamefont
  {Ricouvier}}, \bibinfo {author} {\bibfnamefont {Romain}\ \bibnamefont
  {Pierrat}}, \bibinfo {author} {\bibfnamefont {R{\'e}mi}\ \bibnamefont
  {Carminati}}, \bibinfo {author} {\bibfnamefont {Patrick}\ \bibnamefont
  {Tabeling}}, \ and\ \bibinfo {author} {\bibfnamefont {Pavel}\ \bibnamefont
  {Yazhgur}},\ }\bibfield  {title} {\enquote {\bibinfo {title} {Optimizing
  hyperuniformity in self-assembled bidisperse emulsions},}\ }\href@noop {}
  {\bibfield  {journal} {\bibinfo  {journal} {Phys. Rev. Lett.}\ }\textbf
  {\bibinfo {volume} {119}},\ \bibinfo {pages} {208001} (\bibinfo {year}
  {2017})}\BibitemShut {NoStop}%
\bibitem [{\citenamefont {Jiao}\ \emph {et~al.}(2014)\citenamefont {Jiao},
  \citenamefont {Lau}, \citenamefont {Hatzikirou}, \citenamefont
  {Meyer-Hermann}, \citenamefont {Corbo},\ and\ \citenamefont
  {Torquato}}]{jiao2014avian}%
  \BibitemOpen
  \bibfield  {author} {\bibinfo {author} {\bibfnamefont {Yang}\ \bibnamefont
  {Jiao}}, \bibinfo {author} {\bibfnamefont {Timothy}\ \bibnamefont {Lau}},
  \bibinfo {author} {\bibfnamefont {Haralampos}\ \bibnamefont {Hatzikirou}},
  \bibinfo {author} {\bibfnamefont {Michael}\ \bibnamefont {Meyer-Hermann}},
  \bibinfo {author} {\bibfnamefont {Joseph~C}\ \bibnamefont {Corbo}}, \ and\
  \bibinfo {author} {\bibfnamefont {Salvatore}\ \bibnamefont {Torquato}},\
  }\bibfield  {title} {\enquote {\bibinfo {title} {Avian photoreceptor patterns
  represent a disordered hyperuniform solution to a multiscale packing
  problem},}\ }\href@noop {} {\bibfield  {journal} {\bibinfo  {journal} {Phys.
  Rev. E}\ }\textbf {\bibinfo {volume} {89}},\ \bibinfo {pages} {022721}
  (\bibinfo {year} {2014})}\BibitemShut {NoStop}%
\bibitem [{\citenamefont {Zheng}\ \emph {et~al.}(2020)\citenamefont {Zheng},
  \citenamefont {Li},\ and\ \citenamefont
  {Ciamarra}}]{zheng2020hyperuniformity}%
  \BibitemOpen
  \bibfield  {author} {\bibinfo {author} {\bibfnamefont {Yuanjian}\
  \bibnamefont {Zheng}}, \bibinfo {author} {\bibfnamefont {Yan-Wei}\
  \bibnamefont {Li}}, \ and\ \bibinfo {author} {\bibfnamefont {Massimo~Pica}\
  \bibnamefont {Ciamarra}},\ }\bibfield  {title} {\enquote {\bibinfo {title}
  {Hyperuniformity and density fluctuations at a rigidity transition in a model
  of biological tissues},}\ }\href@noop {} {\bibfield  {journal} {\bibinfo
  {journal} {Soft Matter}\ }\textbf {\bibinfo {volume} {16}},\ \bibinfo {pages}
  {5942--5950} (\bibinfo {year} {2020})}\BibitemShut {NoStop}%
\bibitem [{\citenamefont {Gabrielli}\ \emph {et~al.}(2002)\citenamefont
  {Gabrielli}, \citenamefont {Joyce},\ and\ \citenamefont
  {Labini}}]{gabrielli2002glass}%
  \BibitemOpen
  \bibfield  {author} {\bibinfo {author} {\bibfnamefont {Andrea}\ \bibnamefont
  {Gabrielli}}, \bibinfo {author} {\bibfnamefont {Michael}\ \bibnamefont
  {Joyce}}, \ and\ \bibinfo {author} {\bibfnamefont {Francesco~Sylos}\
  \bibnamefont {Labini}},\ }\bibfield  {title} {\enquote {\bibinfo {title}
  {Glass-like universe: Real-space correlation properties of standard
  cosmological models},}\ }\href@noop {} {\bibfield  {journal} {\bibinfo
  {journal} {Phys. Rev. D}\ }\textbf {\bibinfo {volume} {65}},\ \bibinfo
  {pages} {083523} (\bibinfo {year} {2002})}\BibitemShut {NoStop}%
\bibitem [{\citenamefont {Florescu}\ \emph {et~al.}(2009)\citenamefont
  {Florescu}, \citenamefont {Torquato},\ and\ \citenamefont
  {Steinhardt}}]{florescu2009designer}%
  \BibitemOpen
  \bibfield  {author} {\bibinfo {author} {\bibfnamefont {Marian}\ \bibnamefont
  {Florescu}}, \bibinfo {author} {\bibfnamefont {Salvatore}\ \bibnamefont
  {Torquato}}, \ and\ \bibinfo {author} {\bibfnamefont {Paul~J}\ \bibnamefont
  {Steinhardt}},\ }\bibfield  {title} {\enquote {\bibinfo {title} {Designer
  disordered materials with large, complete photonic band gaps},}\ }\href@noop
  {} {\bibfield  {journal} {\bibinfo  {journal} {Proc. Natl. Acad. Sci. USA}\
  }\textbf {\bibinfo {volume} {106}},\ \bibinfo {pages} {20658--20663}
  (\bibinfo {year} {2009})}\BibitemShut {NoStop}%
\bibitem [{\citenamefont {Man}\ \emph {et~al.}(2013{\natexlab{a}})\citenamefont
  {Man}, \citenamefont {Florescu}, \citenamefont {Williamson}, \citenamefont
  {He}, \citenamefont {Hashemizad}, \citenamefont {Leung}, \citenamefont
  {Liner}, \citenamefont {Torquato}, \citenamefont {Chaikin},\ and\
  \citenamefont {Steinhardt}}]{man2013isotropic}%
  \BibitemOpen
  \bibfield  {author} {\bibinfo {author} {\bibfnamefont {Weining}\ \bibnamefont
  {Man}}, \bibinfo {author} {\bibfnamefont {Marian}\ \bibnamefont {Florescu}},
  \bibinfo {author} {\bibfnamefont {Eric~Paul}\ \bibnamefont {Williamson}},
  \bibinfo {author} {\bibfnamefont {Yingquan}\ \bibnamefont {He}}, \bibinfo
  {author} {\bibfnamefont {Seyed~Reza}\ \bibnamefont {Hashemizad}}, \bibinfo
  {author} {\bibfnamefont {Brian~YC}\ \bibnamefont {Leung}}, \bibinfo {author}
  {\bibfnamefont {Devin~Robert}\ \bibnamefont {Liner}}, \bibinfo {author}
  {\bibfnamefont {Salvatore}\ \bibnamefont {Torquato}}, \bibinfo {author}
  {\bibfnamefont {Paul~M}\ \bibnamefont {Chaikin}}, \ and\ \bibinfo {author}
  {\bibfnamefont {Paul~J}\ \bibnamefont {Steinhardt}},\ }\bibfield  {title}
  {\enquote {\bibinfo {title} {Isotropic band gaps and freeform waveguides
  observed in hyperuniform disordered photonic solids},}\ }\href@noop {}
  {\bibfield  {journal} {\bibinfo  {journal} {Proc. Natl. Acad. Sci. USA}\
  }\textbf {\bibinfo {volume} {110}},\ \bibinfo {pages} {15886--15891}
  (\bibinfo {year} {2013}{\natexlab{a}})}\BibitemShut {NoStop}%
\bibitem [{\citenamefont {Man}\ \emph {et~al.}(2013{\natexlab{b}})\citenamefont
  {Man}, \citenamefont {Florescu}, \citenamefont {Matsuyama}, \citenamefont
  {Yadak}, \citenamefont {Nahal}, \citenamefont {Hashemizad}, \citenamefont
  {Williamson}, \citenamefont {Steinhardt}, \citenamefont {Torquato},\ and\
  \citenamefont {Chaikin}}]{manoe2013}%
  \BibitemOpen
  \bibfield  {author} {\bibinfo {author} {\bibfnamefont {W.}~\bibnamefont
  {Man}}, \bibinfo {author} {\bibfnamefont {M.}~\bibnamefont {Florescu}},
  \bibinfo {author} {\bibfnamefont {K.}~\bibnamefont {Matsuyama}}, \bibinfo
  {author} {\bibfnamefont {P.}~\bibnamefont {Yadak}}, \bibinfo {author}
  {\bibfnamefont {G.}~\bibnamefont {Nahal}}, \bibinfo {author} {\bibfnamefont
  {S.}~\bibnamefont {Hashemizad}}, \bibinfo {author} {\bibfnamefont
  {E.}~\bibnamefont {Williamson}}, \bibinfo {author} {\bibfnamefont
  {P.}~\bibnamefont {Steinhardt}}, \bibinfo {author} {\bibfnamefont
  {S.}~\bibnamefont {Torquato}}, \ and\ \bibinfo {author} {\bibfnamefont
  {P.}~\bibnamefont {Chaikin}},\ }\bibfield  {title} {\enquote {\bibinfo
  {title} {Photonic band gap in isotropic hyperuniform disordered solids with
  low dielectric contrast},}\ }\href@noop {} {\bibfield  {journal} {\bibinfo
  {journal} {Opt. Express}\ }\textbf {\bibinfo {volume} {21}},\ \bibinfo
  {pages} {19972--19981} (\bibinfo {year} {2013}{\natexlab{b}})}\BibitemShut
  {NoStop}%
\bibitem [{\citenamefont {Batten}\ \emph {et~al.}(2008)\citenamefont {Batten},
  \citenamefont {Stillinger},\ and\ \citenamefont {Torquato}}]{transp2008}%
  \BibitemOpen
  \bibfield  {author} {\bibinfo {author} {\bibfnamefont {Robert~D.}\
  \bibnamefont {Batten}}, \bibinfo {author} {\bibfnamefont {Frank~H.}\
  \bibnamefont {Stillinger}}, \ and\ \bibinfo {author} {\bibfnamefont
  {Salvatore}\ \bibnamefont {Torquato}},\ }\bibfield  {title} {\enquote
  {\bibinfo {title} {{Classical disordered ground states: Super-ideal gases and
  stealth and equi-luminous materials}},}\ }\href@noop {} {\bibfield  {journal}
  {\bibinfo  {journal} {J. Appl. Phys.}\ }\textbf {\bibinfo {volume} {104}}
  (\bibinfo {year} {2008})},\ \bibinfo {note} {033504}\BibitemShut {NoStop}%
\bibitem [{\citenamefont {Leseur}\ \emph {et~al.}(2016)\citenamefont {Leseur},
  \citenamefont {Pierrat},\ and\ \citenamefont {Carminati}}]{leseur2016high}%
  \BibitemOpen
  \bibfield  {author} {\bibinfo {author} {\bibfnamefont {Olivier}\ \bibnamefont
  {Leseur}}, \bibinfo {author} {\bibfnamefont {Romain}\ \bibnamefont
  {Pierrat}}, \ and\ \bibinfo {author} {\bibfnamefont {R{\'e}mi}\ \bibnamefont
  {Carminati}},\ }\bibfield  {title} {\enquote {\bibinfo {title} {High-density
  hyperuniform materials can be transparent},}\ }\href@noop {} {\bibfield
  {journal} {\bibinfo  {journal} {Optica}\ }\textbf {\bibinfo {volume} {3}},\
  \bibinfo {pages} {763--767} (\bibinfo {year} {2016})}\BibitemShut {NoStop}%
\bibitem [{\citenamefont {Hexner}\ and\ \citenamefont
  {Levine}(2017)}]{hexner2017noise}%
  \BibitemOpen
  \bibfield  {author} {\bibinfo {author} {\bibfnamefont {Daniel}\ \bibnamefont
  {Hexner}}\ and\ \bibinfo {author} {\bibfnamefont {Dov}\ \bibnamefont
  {Levine}},\ }\bibfield  {title} {\enquote {\bibinfo {title} {Noise,
  diffusion, and hyperuniformity},}\ }\href@noop {} {\bibfield  {journal}
  {\bibinfo  {journal} {Phys. Rev. Lett.}\ }\textbf {\bibinfo {volume} {118}},\
  \bibinfo {pages} {020601} (\bibinfo {year} {2017})}\BibitemShut {NoStop}%
\bibitem [{\citenamefont {Lei}\ and\ \citenamefont
  {Ni}(2019)}]{lei2019hydrodynamics}%
  \BibitemOpen
  \bibfield  {author} {\bibinfo {author} {\bibfnamefont {Qun-Li}\ \bibnamefont
  {Lei}}\ and\ \bibinfo {author} {\bibfnamefont {Ran}\ \bibnamefont {Ni}},\
  }\bibfield  {title} {\enquote {\bibinfo {title} {Hydrodynamics of
  random-organizing hyperuniform fluids},}\ }\href@noop {} {\bibfield
  {journal} {\bibinfo  {journal} {Proc. Natl. Acad. Sci. USA}\ }\textbf
  {\bibinfo {volume} {116}},\ \bibinfo {pages} {22983--22989} (\bibinfo {year}
  {2019})}\BibitemShut {NoStop}%
\bibitem [{\citenamefont {Mitra}\ \emph {et~al.}(2021)\citenamefont {Mitra},
  \citenamefont {Parmar}, \citenamefont {Leishangthem}, \citenamefont
  {Sastry},\ and\ \citenamefont {Foffi}}]{Mitra_2021}%
  \BibitemOpen
  \bibfield  {author} {\bibinfo {author} {\bibfnamefont {Saheli}\ \bibnamefont
  {Mitra}}, \bibinfo {author} {\bibfnamefont {Anshul D~S}\ \bibnamefont
  {Parmar}}, \bibinfo {author} {\bibfnamefont {Premkumar}\ \bibnamefont
  {Leishangthem}}, \bibinfo {author} {\bibfnamefont {Srikanth}\ \bibnamefont
  {Sastry}}, \ and\ \bibinfo {author} {\bibfnamefont {Giuseppe}\ \bibnamefont
  {Foffi}},\ }\bibfield  {title} {\enquote {\bibinfo {title} {Hyperuniformity
  in cyclically driven glasses},}\ }\href@noop {} {\bibfield  {journal}
  {\bibinfo  {journal} {J. Stat. Mech.: Theory Exp}\ }\textbf {\bibinfo
  {volume} {2021}},\ \bibinfo {pages} {033203} (\bibinfo {year}
  {2021})}\BibitemShut {NoStop}%
\bibitem [{\citenamefont {Hexner}\ and\ \citenamefont
  {Levine}(2015)}]{hexner2015hyperuniformity}%
  \BibitemOpen
  \bibfield  {author} {\bibinfo {author} {\bibfnamefont {Daniel}\ \bibnamefont
  {Hexner}}\ and\ \bibinfo {author} {\bibfnamefont {Dov}\ \bibnamefont
  {Levine}},\ }\bibfield  {title} {\enquote {\bibinfo {title} {Hyperuniformity
  of critical absorbing states},}\ }\href@noop {} {\bibfield  {journal}
  {\bibinfo  {journal} {Phys. Rev. Lett.}\ }\textbf {\bibinfo {volume} {114}},\
  \bibinfo {pages} {110602} (\bibinfo {year} {2015})}\BibitemShut {NoStop}%
\bibitem [{\citenamefont {Tjhung}\ and\ \citenamefont
  {Berthier}(2015)}]{tjhung2015hyperuniform}%
  \BibitemOpen
  \bibfield  {author} {\bibinfo {author} {\bibfnamefont {Elsen}\ \bibnamefont
  {Tjhung}}\ and\ \bibinfo {author} {\bibfnamefont {Ludovic}\ \bibnamefont
  {Berthier}},\ }\bibfield  {title} {\enquote {\bibinfo {title} {Hyperuniform
  density fluctuations and diverging dynamic correlations in periodically
  driven colloidal suspensions},}\ }\href@noop {} {\bibfield  {journal}
  {\bibinfo  {journal} {Phys. Rev. Lett.}\ }\textbf {\bibinfo {volume} {114}},\
  \bibinfo {pages} {148301} (\bibinfo {year} {2015})}\BibitemShut {NoStop}%
\bibitem [{\citenamefont {Lei}\ \emph {et~al.}(2019)\citenamefont {Lei},
  \citenamefont {Ciamarra},\ and\ \citenamefont {Ni}}]{lei2019nonequilibrium}%
  \BibitemOpen
  \bibfield  {author} {\bibinfo {author} {\bibfnamefont {Qun-Li}\ \bibnamefont
  {Lei}}, \bibinfo {author} {\bibfnamefont {Massimo~Pica}\ \bibnamefont
  {Ciamarra}}, \ and\ \bibinfo {author} {\bibfnamefont {Ran}\ \bibnamefont
  {Ni}},\ }\bibfield  {title} {\enquote {\bibinfo {title} {Nonequilibrium
  strongly hyperuniform fluids of circle active particles with large local
  density fluctuations},}\ }\href@noop {} {\bibfield  {journal} {\bibinfo
  {journal} {Sci. Adv.}\ }\textbf {\bibinfo {volume} {5}},\ \bibinfo {pages}
  {eaau7423} (\bibinfo {year} {2019})}\BibitemShut {NoStop}%
\bibitem [{\citenamefont {Wang}\ \emph {et~al.}(2018)\citenamefont {Wang},
  \citenamefont {Schwarz},\ and\ \citenamefont
  {Paulsen}}]{wang2018hyperuniformity}%
  \BibitemOpen
  \bibfield  {author} {\bibinfo {author} {\bibfnamefont {Jikai}\ \bibnamefont
  {Wang}}, \bibinfo {author} {\bibfnamefont {Jennifer~M}\ \bibnamefont
  {Schwarz}}, \ and\ \bibinfo {author} {\bibfnamefont {Joseph~D}\ \bibnamefont
  {Paulsen}},\ }\bibfield  {title} {\enquote {\bibinfo {title} {Hyperuniformity
  with no fine tuning in sheared sedimenting suspensions},}\ }\href@noop {}
  {\bibfield  {journal} {\bibinfo  {journal} {Nat. Commun.}\ }\textbf {\bibinfo
  {volume} {9}},\ \bibinfo {pages} {1--7} (\bibinfo {year} {2018})}\BibitemShut
  {NoStop}%
\bibitem [{\citenamefont {Oppenheimer}\ \emph {et~al.}(2022)\citenamefont
  {Oppenheimer}, \citenamefont {Stein}, \citenamefont {Zion},\ and\
  \citenamefont {Shelley}}]{oppenheimer2022hyperuniformity}%
  \BibitemOpen
  \bibfield  {author} {\bibinfo {author} {\bibfnamefont {Naomi}\ \bibnamefont
  {Oppenheimer}}, \bibinfo {author} {\bibfnamefont {David~B}\ \bibnamefont
  {Stein}}, \bibinfo {author} {\bibfnamefont {Matan Yah~Ben}\ \bibnamefont
  {Zion}}, \ and\ \bibinfo {author} {\bibfnamefont {Michael~J}\ \bibnamefont
  {Shelley}},\ }\bibfield  {title} {\enquote {\bibinfo {title} {Hyperuniformity
  and phase enrichment in vortex and rotor assemblies},}\ }\href@noop {}
  {\bibfield  {journal} {\bibinfo  {journal} {Nat. Commun.}\ }\textbf {\bibinfo
  {volume} {13}},\ \bibinfo {pages} {804} (\bibinfo {year} {2022})}\BibitemShut
  {NoStop}%
\bibitem [{\citenamefont {Zhang}\ and\ \citenamefont
  {Snezhko}(2022)}]{zhang2022hyperuniform}%
  \BibitemOpen
  \bibfield  {author} {\bibinfo {author} {\bibfnamefont {Bo}~\bibnamefont
  {Zhang}}\ and\ \bibinfo {author} {\bibfnamefont {Alexey}\ \bibnamefont
  {Snezhko}},\ }\bibfield  {title} {\enquote {\bibinfo {title} {Hyperuniform
  active chiral fluids with tunable internal structure},}\ }\href@noop {}
  {\bibfield  {journal} {\bibinfo  {journal} {Phys. Rev. Lett.}\ }\textbf
  {\bibinfo {volume} {128}},\ \bibinfo {pages} {218002} (\bibinfo {year}
  {2022})}\BibitemShut {NoStop}%
\bibitem [{\citenamefont {Huang}\ \emph {et~al.}(2021)\citenamefont {Huang},
  \citenamefont {Hu}, \citenamefont {Yang}, \citenamefont {Liu},\ and\
  \citenamefont {Zhang}}]{huang2021circular}%
  \BibitemOpen
  \bibfield  {author} {\bibinfo {author} {\bibfnamefont {Mingji}\ \bibnamefont
  {Huang}}, \bibinfo {author} {\bibfnamefont {Wensi}\ \bibnamefont {Hu}},
  \bibinfo {author} {\bibfnamefont {Siyuan}\ \bibnamefont {Yang}}, \bibinfo
  {author} {\bibfnamefont {Quan-Xing}\ \bibnamefont {Liu}}, \ and\ \bibinfo
  {author} {\bibfnamefont {HP}~\bibnamefont {Zhang}},\ }\bibfield  {title}
  {\enquote {\bibinfo {title} {Circular swimming motility and disordered
  hyperuniform state in an algae system},}\ }\href@noop {} {\bibfield
  {journal} {\bibinfo  {journal} {Proc. Natl. Acad. Sci. USA}\ }\textbf
  {\bibinfo {volume} {118}},\ \bibinfo {pages} {e2100493118} (\bibinfo {year}
  {2021})}\BibitemShut {NoStop}%
\bibitem [{\citenamefont {Wilken}\ \emph {et~al.}(2020)\citenamefont {Wilken},
  \citenamefont {Guerra}, \citenamefont {Pine},\ and\ \citenamefont
  {Chaikin}}]{chaikinprl2020}%
  \BibitemOpen
  \bibfield  {author} {\bibinfo {author} {\bibfnamefont {Sam}\ \bibnamefont
  {Wilken}}, \bibinfo {author} {\bibfnamefont {Rodrigo~E.}\ \bibnamefont
  {Guerra}}, \bibinfo {author} {\bibfnamefont {David~J.}\ \bibnamefont {Pine}},
  \ and\ \bibinfo {author} {\bibfnamefont {Paul~M.}\ \bibnamefont {Chaikin}},\
  }\bibfield  {title} {\enquote {\bibinfo {title} {Hyperuniform structures
  formed by shearing colloidal suspensions},}\ }\href {\doibase
  10.1103/PhysRevLett.125.148001} {\bibfield  {journal} {\bibinfo  {journal}
  {Phys. Rev. Lett.}\ }\textbf {\bibinfo {volume} {125}},\ \bibinfo {pages}
  {148001} (\bibinfo {year} {2020})}\BibitemShut {NoStop}%
\bibitem [{\citenamefont {Weijs}\ \emph {et~al.}(2015)\citenamefont {Weijs},
  \citenamefont {Jeanneret}, \citenamefont {Dreyfus},\ and\ \citenamefont
  {Bartolo}}]{weijsprl2015}%
  \BibitemOpen
  \bibfield  {author} {\bibinfo {author} {\bibfnamefont {Joost~H.}\
  \bibnamefont {Weijs}}, \bibinfo {author} {\bibfnamefont {Rapha\"el}\
  \bibnamefont {Jeanneret}}, \bibinfo {author} {\bibfnamefont {R\'emi}\
  \bibnamefont {Dreyfus}}, \ and\ \bibinfo {author} {\bibfnamefont {Denis}\
  \bibnamefont {Bartolo}},\ }\bibfield  {title} {\enquote {\bibinfo {title}
  {Emergent hyperuniformity in periodically driven emulsions},}\ }\href
  {\doibase 10.1103/PhysRevLett.115.108301} {\bibfield  {journal} {\bibinfo
  {journal} {Phys. Rev. Lett.}\ }\textbf {\bibinfo {volume} {115}},\ \bibinfo
  {pages} {108301} (\bibinfo {year} {2015})}\BibitemShut {NoStop}%
\bibitem [{\citenamefont {Lei}\ \emph {et~al.}(2021)\citenamefont {Lei},
  \citenamefont {Hu},\ and\ \citenamefont {Ni}}]{lei2021barrier}%
  \BibitemOpen
  \bibfield  {author} {\bibinfo {author} {\bibfnamefont {Qun-Li}\ \bibnamefont
  {Lei}}, \bibinfo {author} {\bibfnamefont {Hao}\ \bibnamefont {Hu}}, \ and\
  \bibinfo {author} {\bibfnamefont {Ran}\ \bibnamefont {Ni}},\ }\bibfield
  {title} {\enquote {\bibinfo {title} {Barrier-controlled nonequilibrium
  criticality in reactive particle systems},}\ }\href@noop {} {\bibfield
  {journal} {\bibinfo  {journal} {Phys. Rev. E}\ }\textbf {\bibinfo {volume}
  {103}},\ \bibinfo {pages} {052607} (\bibinfo {year} {2021})}\BibitemShut
  {NoStop}%
\bibitem [{\citenamefont {Villa~Mart{\'\i}n}\ \emph {et~al.}(2015)\citenamefont
  {Villa~Mart{\'\i}n}, \citenamefont {Bonachela}, \citenamefont {Levin},\ and\
  \citenamefont {Mu{\~n}oz}}]{villa2015eluding}%
  \BibitemOpen
  \bibfield  {author} {\bibinfo {author} {\bibfnamefont {Paula}\ \bibnamefont
  {Villa~Mart{\'\i}n}}, \bibinfo {author} {\bibfnamefont {Juan~A}\ \bibnamefont
  {Bonachela}}, \bibinfo {author} {\bibfnamefont {Simon~A}\ \bibnamefont
  {Levin}}, \ and\ \bibinfo {author} {\bibfnamefont {Miguel~A}\ \bibnamefont
  {Mu{\~n}oz}},\ }\bibfield  {title} {\enquote {\bibinfo {title} {Eluding
  catastrophic shifts},}\ }\href@noop {} {\bibfield  {journal} {\bibinfo
  {journal} {Proc. Natl. Acad. Sci. USA}\ }\textbf {\bibinfo {volume} {112}},\
  \bibinfo {pages} {E1828--E1836} (\bibinfo {year} {2015})}\BibitemShut
  {NoStop}%
\bibitem [{\citenamefont {Kawasaki}\ and\ \citenamefont
  {Berthier}(2016)}]{ludopre2016}%
  \BibitemOpen
  \bibfield  {author} {\bibinfo {author} {\bibfnamefont {Takeshi}\ \bibnamefont
  {Kawasaki}}\ and\ \bibinfo {author} {\bibfnamefont {Ludovic}\ \bibnamefont
  {Berthier}},\ }\bibfield  {title} {\enquote {\bibinfo {title} {Macroscopic
  yielding in jammed solids is accompanied by a nonequilibrium first-order
  transition in particle trajectories},}\ }\href@noop {} {\bibfield  {journal}
  {\bibinfo  {journal} {Phys. Rev. E}\ }\textbf {\bibinfo {volume} {94}},\
  \bibinfo {pages} {022615} (\bibinfo {year} {2016})}\BibitemShut {NoStop}%
\bibitem [{\citenamefont {Tjhung}\ and\ \citenamefont
  {Berthier}(2017)}]{ludopre2017}%
  \BibitemOpen
  \bibfield  {author} {\bibinfo {author} {\bibfnamefont {Elsen}\ \bibnamefont
  {Tjhung}}\ and\ \bibinfo {author} {\bibfnamefont {Ludovic}\ \bibnamefont
  {Berthier}},\ }\bibfield  {title} {\enquote {\bibinfo {title} {Discontinuous
  fluidization transition in time-correlated assemblies of actively deforming
  particles},}\ }\href@noop {} {\bibfield  {journal} {\bibinfo  {journal}
  {Phys. Rev. E}\ }\textbf {\bibinfo {volume} {96}},\ \bibinfo {pages} {050601}
  (\bibinfo {year} {2017})}\BibitemShut {NoStop}%
\bibitem [{\citenamefont {Ni}\ \emph {et~al.}(2010)\citenamefont {Ni},
  \citenamefont {Belli}, \citenamefont {van Roij},\ and\ \citenamefont
  {Dijkstra}}]{niprl2010}%
  \BibitemOpen
  \bibfield  {author} {\bibinfo {author} {\bibfnamefont {Ran}\ \bibnamefont
  {Ni}}, \bibinfo {author} {\bibfnamefont {Simone}\ \bibnamefont {Belli}},
  \bibinfo {author} {\bibfnamefont {Ren\'e}\ \bibnamefont {van Roij}}, \ and\
  \bibinfo {author} {\bibfnamefont {Marjolein}\ \bibnamefont {Dijkstra}},\
  }\bibfield  {title} {\enquote {\bibinfo {title} {Glassy dynamics, spinodal
  fluctuations, and the kinetic limit of nucleation in suspensions of colloidal
  hard rods},}\ }\href@noop {} {\bibfield  {journal} {\bibinfo  {journal}
  {Phys. Rev. Lett.}\ }\textbf {\bibinfo {volume} {105}},\ \bibinfo {pages}
  {088302} (\bibinfo {year} {2010})}\BibitemShut {NoStop}%
\end{thebibliography}%

\end{document}